\documentclass[twocolumn,aps,prl,showpacs,superscriptaddress]{revtex4}
\usepackage{amssymb,amsmath}
\usepackage{graphicx}

\newcommand{\units}[1]{\,\ensuremath{\mathrm{#1}}}

\newcommand{\Soo}{\ensuremath{\mathrm{S_{11}}}}
\newcommand{\Sot}{\ensuremath{\mathrm{S_{02}}}}
\newcommand{\To}{\ensuremath{\mathrm{T_{0}}}}
\newcommand{\Tp}{\ensuremath{\mathrm{T_{+}}}}
\newcommand{\Tm}{\ensuremath{\mathrm{T_{-}}}}
\newcommand{\Too}{\ensuremath{\mathrm{T_{11}}}}

\begin{document}

\title{Single-shot measurement of triplet-singlet relaxation in a Si/SiGe double quantum dot}

\newcommand{\uwaff}{\affiliation{University of Wisconsin-Madison, Madison, WI 53706}}
\newcommand{\delftaff}{\affiliation{Kavli Institute of Nanoscience, TU Delft, Lorentzweg 1, 2628 CJ Delft, The Netherlands}}

\author{J. R. Prance}\uwaff
\author{Zhan Shi}\uwaff
\author{C. B. Simmons}\uwaff
\author{D. E. Savage}\uwaff
\author{M. G. Lagally}\uwaff
\author{L. R. Schreiber}\delftaff
\author{L. M. K. Vandersypen}\delftaff
\author{Mark Friesen}\uwaff
\author{Robert Joynt}\uwaff
\author{S. N. Coppersmith}\uwaff
\author{M. A. Eriksson}\uwaff


\begin{abstract}
We investigate the lifetime of two-electron spin states in a few-electron Si/SiGe double dot. At the transition between the (1,1) and (0,2) charge occupations, Pauli spin blockade provides a readout mechanism for the spin state. We use the statistics of repeated single-shot measurements to extract the lifetimes of multiple states simultaneously. At zero magnetic field, we find that all three triplet states have equal lifetimes, as expected, and this time is $\sim10\units{ms}$. At non-zero field, the \To\ lifetime is unchanged, whereas the \Tm\ lifetime increases monotonically with field, reaching $3$ seconds at $1\units{T}$.
\end{abstract}

\maketitle

The lifetimes of single electron spins in silicon have recently been measured to be as long as seconds in Si nanodevices, including gated quantum dots and donors \cite{Xiao:2010p096801,Morello:2010p687,Simmons:2011p156804,Hayes:2009preprint}, a promising step towards silicon spin qubits.  Two-electron singlet-triplet states in a double dot can also be used as qubits \cite{Levy:2002p1446,Petta:2005p2180,Foletti:2009p903}, with the advantages that gating operations can be fast and that readout depends on the singlet-triplet energy splitting, which can be much larger than the single spin Zeeman energy at low magnetic fields. The lifetimes of singlet and triplet states have been measured in GaAs double dots and were found to depend on magnetic field, falling to $<30\units{\mu s}$ at zero field \cite{Johnson:2005p925,Petta:2005p161301}.  In silicon, neither single-shot readout of the singlet-triplet qubit states, nor measurement of their lifetimes has been achieved up until now.

Here we report measurements of the lifetimes of singlet and triplet states in a Si/SiGe double quantum dot at magnetic fields from $1\units{T}$ to $0\units{T}$ obtained using single-shot read-out. Using pulsed gate voltages, we repeatedly alternate the charge detuning so that it first favors the (1,1) charge state (one electron in each dot) and then the (0,2) charge state (two electrons in one of the dots.) Because of Pauli spin blockade, charge transitions to (0,2) will only occur when the spin state is a singlet. We perform hundreds of thousands of such cycles and measure the presence or absence of charge transitions using real-time charge sensing.  By analyzing the statistics of such data, we characterize multiple relaxation processes simultaneously, in contrast to time-averaged measurements, which are only sensitive to the rate-limiting process.  At zero magnetic field the triplet and singlet state lifetimes are between 5 and 25 ms, lifetimes that exceed those measured in GaAs by over two orders of magnitude. As magnetic field increases, the lifetime of the \To\ remains essentially constant, whereas the lifetime of the \Tm\ increases dramatically, reaching 3 seconds at $B_{||}=1$ T.  These long times are expected because of the small hyperfine coupling and spin-orbit interaction in Si quantum dots.

\begin{figure}\centering
\includegraphics{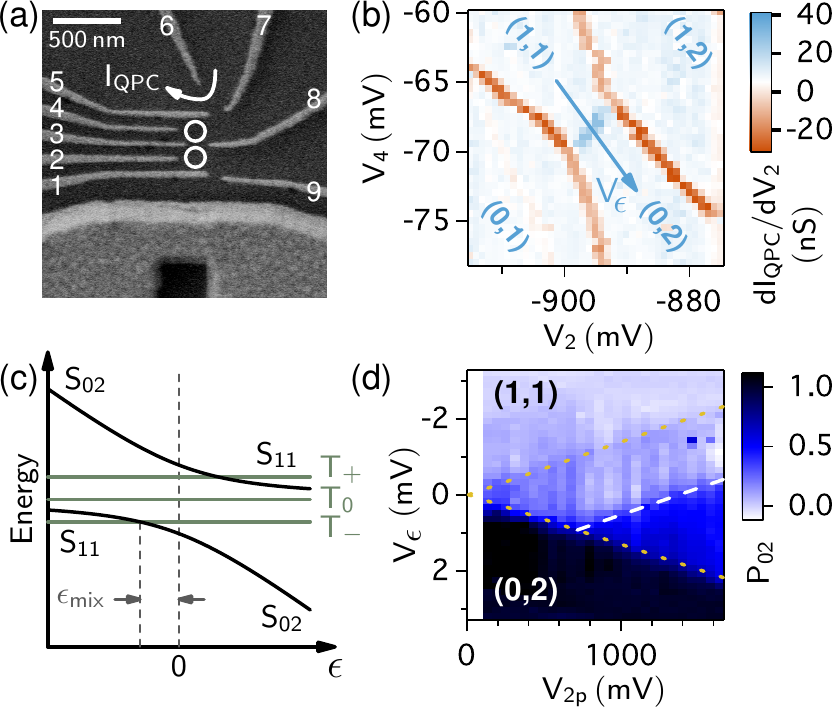}
\caption{\label{fig1} (a) SEM image of a device identical to the one used. Quantum dots are formed at the approximate locations of the two circles. Charge sensing is performed by monitoring the current $I_{QPC}$ through a nearby point-contact. (b) Charge stability diagram of the double dot showing the detuning voltage $V_\epsilon$. (c) Energies of two-electron states as a function of detuning energy $\epsilon$. \Tp, \To\ and \Tm\ are the (1,1) triplets; the (0,2) triplets are higher in energy. The (1,1) and (0,2) singlets \Soo\ and \Sot\ are coupled by spin-preserving, inter-dot tunneling. A magnetic field separates the triplet energies by $E_z = g\mu_B B$. (d) Time-averaged occupation of the (0,2) charge state $P_{02}$ at $B_{||} = 0$ with $5\units{kHz}$ square pulses applied along $V_\epsilon$: pulse amplitudes on gates 2 and 4 are related by $V_{4p} = -0.35 V_{2p}$. The pulses drive (1,1)-(0,2) transitions within the dotted triangle. The suppression of $P_{02}$ above the dashed line shows where (1,1) to (0,2) tunneling is suppressed by spin blockade.}
\end{figure}

The device is fabricated on a phosphorus-doped $\mathrm{Si/Si_{0.7}Ge_{0.3}}$ heterostructure with a strained Si quantum well approximately $75\units{nm}$ below the surface. Palladium surface gates labelled 1-9 in Fig.~\ref{fig1}(a) are used to form the double-dot confinement potential \cite{Simmons:2009p3234}. A thick RF antenna (Ti/Au, $5\units{nm}$/$305\units{nm}$) is also present near the dot gates, but is unused in this experiment. All gates are connected to room temperature voltage sources via cold RC filters, which are at the measurement base temperature of $\approx 15\units{mK}$. Gates 2 and 4 are also AC coupled to coaxial lines, allowing them to be pulsed at frequencies between $100\units{Hz}$ and $1\units{GHz}$. There is an attenuation of $\approx 50\units{dB}$ between each gate and the pulse source (a Tektronix AFG3252.) Current through the device is measured with a room-temperature current preamplifier with a bandwidth $\approx 1\units{kHz}$.

Figure~\ref{fig1}(b) shows a charge stability diagram in which the absolute occupation of the dots was found by emptying both dots and then counting electrons back in. Fig.~\ref{fig1}(c) shows the predicted energies of the two-electron states near the (1,1)-(0,2) transition as a function of detuning energy $\epsilon$, where the transition is at $\epsilon = 0$ \cite{Hanson:2007p1217}. The detuning energy is controlled by varying the voltages on gates 2 and 4 along $V_\epsilon$, shown in Fig.~\ref{fig1}(b). The inter-dot tunnel coupling $t_c$ was measured by determining where the \Soo\ and \Tm\ states cross at finite $B_{||}$. This is shown as $\epsilon_{mix}$ in Fig.~\ref{fig1}(c), and depends on both $B_{||}$ and the curvature of the avoided singlet crossing. Using this approach \cite{Petta:2005p2180}, we find $t_c = 2.8 \pm 0.3\units{\mu eV}$ ($677 \pm 73\units{MHz}$.)

To measure the spin of a (1,1) state we pulse the system into a spin blockaded configuration \cite{Shaji:2008p540,Borselli:2011p063109,Lai:2010preprint}, where the ground state of the system is \Sot\ and the (0,2) triplet states are higher in energy than all of the (1,1) triplets: \Tm, \To\ and \Tp. We characterize the parameters needed to reach this configuration by detecting spin blockade in the time-averaged measurement shown in Fig.~\ref{fig1}(d). Square pulses at $5\units{kHz}$ are applied along $V_\epsilon$. The color scale in Fig.~\ref{fig1}(d) shows the time-averaged probability $P_{02}$ of finding the system in (0,2) as a function of pulse amplitude and offset along $V_\epsilon$. When the pulse crosses the (1,1)-(0,2) transition, tunneling between charge states results in $0<P_{02}<1$. The region where this occurs is bounded by the dotted triangle in Fig.~\ref{fig1}(d). Spin blockade occurs in the part of the pulse triangle that is above the dashed white line in Fig.~\ref{fig1}(d). Here we see $0 < P_{02} < 0.5$, because the system is residing in (1,1) the majority of the time.

Spin blockade does not occur below the white dashed line in Fig.~\ref{fig1}(d), resulting in $P_{02}\approx 0.5$. In this region the pulse amplitude exceeds the (0,2) singlet-triplet splitting energy $E_{ST}$, and the pulse offset is such that the (0,2) triplet states have lower energy than the (1,1) triplets. From the size of the blockaded region, and the conversion from detuning voltage $V_\epsilon$ to detuning energy $\epsilon$ ($\Delta \epsilon = \Delta V_{\epsilon}\cdot0.0676\units{eV/V}$, see supplemental material below for details), we find $E_{ST}=124\pm4 \units{\mu eV}$.

\begin{figure}
\includegraphics{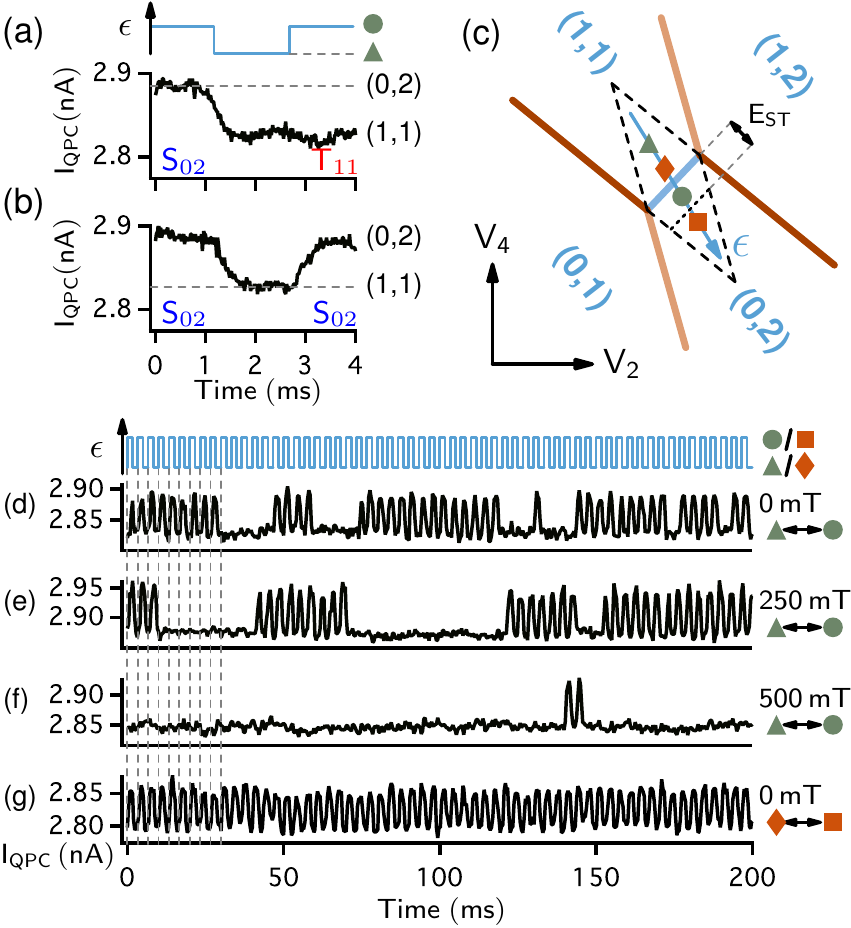}
\caption{\label{fig2} Single-shot initialization and readout of singlet and triplet states. (a),(b) Real-time measurements of $I_{QPC}$ as the system is initialized to \Soo\ then read out $1.7\units{ms}$ later. We identify the final state in (a) as one of the (1,1) triplets (\Too) because the (1,1) charge state survives for over $1\units{ms}$ during the readout. In (b) a singlet is identified because the system tunnels quickly back to (0,2) during the readout. (c) Schematic stability diagram. The points marked are the four detuning values used in the measurements. At $B_{||} > 0$, $E_{ST}$ is decreased by $g\mu_B B_{||}$. The pulse is offset to keep the circle inside the blockaded region without changing the separation of the circle and triangle points. Dashed triangles bound the region where (1,1)-(0,2) transitions occur primarily by inter-dot tunneling. (d)-(g) Pulses repeatedly switch the ground state between (1,1) and (0,2) at $300\units{Hz}$. In (d)-(f) the system is often blockaded in a (1,1) triplet. With increasing magnetic field from (d) to (f), the durations of blockade increase significantly. In (g), the pulse reaches into (0,2) far enough to exceed $E_{ST}$, and tunneling from (1,1) to (0,2) occurs freely for all spin states.}
\end{figure}

Figure~\ref{fig2}(a) and (b) show single-shot initialization and readout of (1,1) singlet and triplet states using real-time measurement of the charge state while pulsing across the (1,1)-(0,2) transition. The system is initialized by starting from the ground state \Sot\ at $0 < \epsilon < E_{ST}$. The occupation of \Sot\ is verified by measuring the charge state: \Sot\ is the only (0,2) state accessible at this detuning. We then pulse to $\epsilon < 0$ to transfer the prepared \Sot\ to the (1,1) singlet \Soo. To measure the (1,1) spin state at some later time, we pulse back to $0 < \epsilon < E_{ST}$ where a singlet can tunnel quickly to (0,2) but the triplets cannot. The measurements are performed using detuning pulses with two levels that are at the positions of the filled triangle and circle in Fig.~\ref{fig2}(c), which correspond to detuning energies of $\epsilon \approx -160\units{\mu eV}$ and $60\units{\mu eV}$ respectively at $B_{||} = 0$.

We measure the lifetimes of the (1,1) singlet and triplet states by detecting the spin state as we repeatedly pulse back and forth across the (1,1)-(0,2) transition at a frequency of $300\units{Hz}$. Fig.~\ref{fig2}(d)-(f) show real-time measurements of the charge state as the pulses are applied. In this regime spin blockade is active and the system switches randomly between free shuttling of a singlet state and blockade of a (1,1) triplet state. The typical length of time spent in a blockaded triplet increases dramatically as $B_{||}$ increases.  Fig~\ref{fig2}(g) is a control, demonstrating that charge shuttles freely in both directions when the pulse is offset to reach outside the spin-blockade regime.

\begin{figure}
\includegraphics{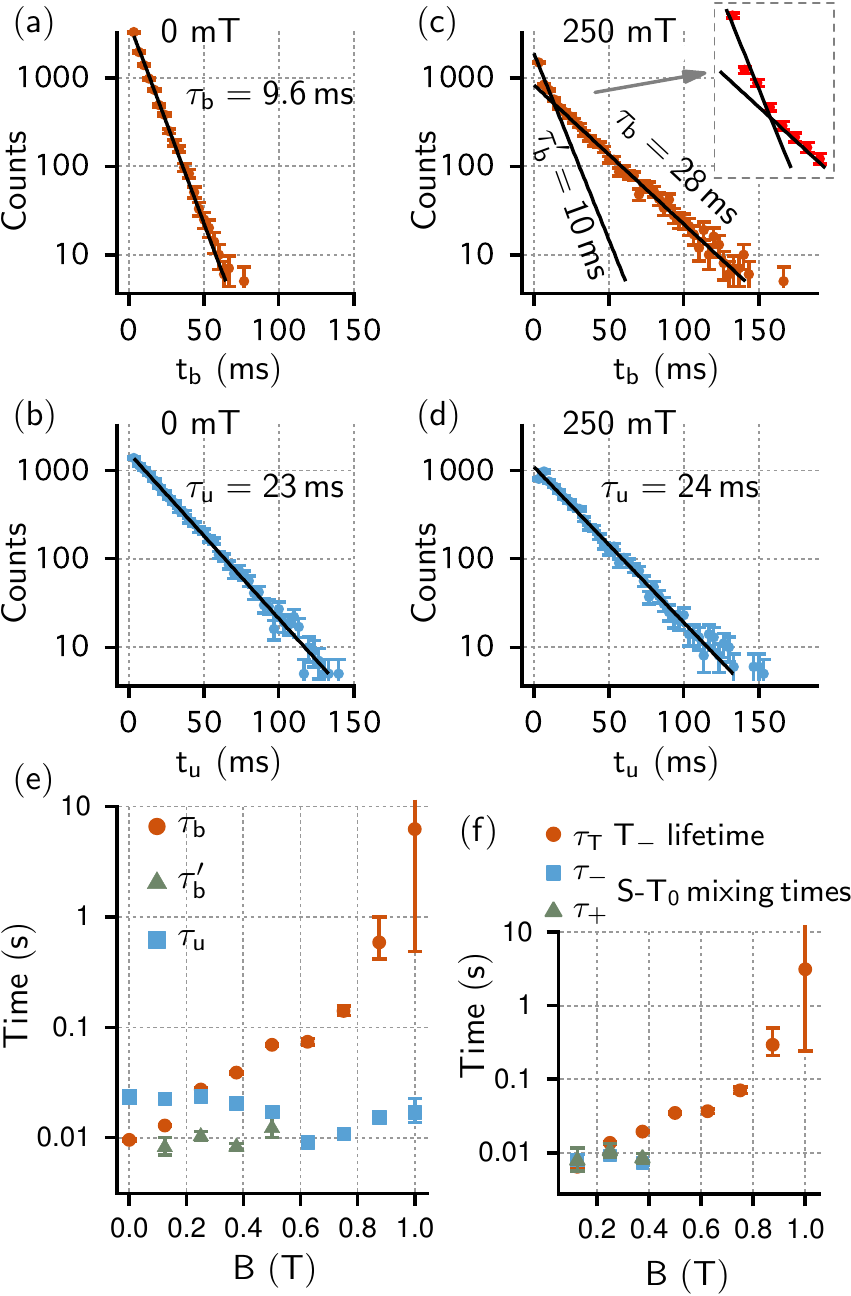}
\caption{\label{fig3} Extracting lifetimes from real-time measurements. (a) Histogram of the number of times that the system is blockaded for a time $t_b$ in many measurements such as Fig.~\ref{fig2}(d). The binning resolution is the pulse period. The solid line is an exponential fit yielding a lifetime for the blockaded configuration of $\tau_b = 9.6\units{ms}$. (b) Histogram of un-blockaded times ($t_u$) for the same data as (a). An exponential fit yields a lifetime for the un-blockaded configuration of $\tau_u = 23\units{ms}$. (c), (d) Histograms of $t_u$ and $t_b$ at $B_{||} = 250\units{mT}$. There are two decay rates describing blockade: at small $t_b$ there is an decay similar to that at zero field ($\tau_b^\prime=10\units{ms}$). At long $t_b$ a slower decay dominates ($\tau_b=28\mathrm{ms}$). We interpret the shorter lifetime as that of the \To, and the longer lifetime as that of the \Tm. (e) Fitted lifetimes as a function of magnetic field. The lifetime of blockade due to \Tm\ states ($\tau_b$) increases with field, while the contribution from \To\ and \Soo\ states ($\tau_b^\prime$ and $\tau_u$) is field independent. (f) \Tm\ lifetime $\tau_T$, and \Soo-\To\ mixing rate at positive (negative) detuning $\tau_+$ ($\tau_-$).}
\end{figure}

To determine the lifetimes of the states at $B_{||} = 0$ we plot in Fig.~\ref{fig3}(a) and (b) the number of times that blockaded periods of duration $t_b$ and un-blockaded periods of duration $t_u$ are observed in $6.4$ minutes of data ($115,200$ pulse periods). The histograms are very well fit by exponential decays, and fits to the two distributions give lifetimes of $\tau_b=9.6\pm0.2 \units{ms}$ for the blockaded configuration and of $\tau_u=23\pm3 \units{ms}$ for the un-blockaded configuration. This shows that the lifetimes of the singlet and triplet states are $\sim 10\units{ms}$.

The $B_{||} = 0$ lifetimes are two orders of magnitude longer than have been seen in comparable low-field measurements of GaAs quantum dots \cite{Johnson:2005p925,Petta:2005p161301}. We suggest that this is due to the small hyperfine coupling in natural silicon, arising from the high abundance of zero-spin nuclei. At $B_{||}=0$, the (1,1) triplets are degenerate and separated from \Soo\ by an energy $J(\epsilon) \approx t_c^2/\epsilon$. We expect singlet-triplet mixing to be driven by a small magnetic field difference between the two dots, resulting from the contact-hyperfine interaction with nuclear spins \cite{Coish:2005p125337,Taylor:2007p464,Assali:2011p165301}. Predictions for the hyperfine coupling of (1,1) spin states are $h \sim 3\units{neV}$ in silicon \cite{Assali:2011p165301}, compared to measured values of $h \sim 50\units{neV}$ in GaAs \cite{Johnson:2005p925,Koppens:2005p717}.  The expected coupling is small enough that, in our measurements, it would be exceeded by the exchange splitting $J$. Given $t_c$ and the pulse amplitude, hyperfine induced singlet-triplet mixing should be suppressed by a factor of $(1 + (J/h)^2) \sim 500$, compared to the maximum mixing rate when $J \ll h$.

The values $\tau_u$ and $\tau_b$ show the time scale of singlet-triplet mixing, but they do not directly correspond to mixing times in any static configuration of the system. This is because the pulses continuously switch between two configurations, one at $\epsilon < 0$ and one at $\epsilon > 0$. The singlet-triplet mixing times may be different in the two configurations, and at $\epsilon > 0$ there are also fast, one-way transitions from \Soo\ to \Sot. We relate the measured values of $\tau_b$ and $\tau_u$ to singlet-triplet mixing times in the two configurations of the system by using rate equations to model state occupations during a single pulse cycle. The inputs to the model are two times; one time $\tau_-$ is the mixing time when the ground state is \Soo\ during the $\epsilon<0$ half of the pulse, and the other time $\tau_+$ is the mixing time when the ground state is \Sot\ during the $\epsilon>0$ half of the pulse. Tunneling between \Soo\ and \Sot\ is assumed to be instantaneous. Mixing during the pulse transitions is ignored because the period of the pulse is $10^5$ times larger than the pulse rise time. We solve for $\tau_+$ and $\tau_-$ by numerical optimization of the model to match the measured values of $\tau_u$ and $\tau_b$ (see supplemental material below for details.) We find $\tau_- = 24.5 \pm 3 \units{ms}$ and $\tau_+ = 5.8\pm 0.3 \units{ms}$.  We attribute the difference between $\tau_+$ and $\tau_-$ to a difference in $t_c$ between the two halves of each pulse cycle.

As $B_{||}$ increases from 0 T, we observe a qualitative change in the spin dynamics: the statistics of the blockaded durations show two separate decay constants. As shown in Fig.~\ref{fig3}(c) and (e), there are short blockaded periods whose typical lifetime $\tau_b^\prime$ is field independent, and there are longer blockaded periods whose lifetime $\tau_b$ increases with field. The two lifetimes arise because the system can be blockaded if it is in either a \To\ or a \Tm\ state, and the \Tm\ has a field dependent energy, whereas the \To\ does not. The \Tp\ state does not play a role at $B_{||} > 0$ because its higher energy means that it is rarely populated. Combined with statistics of un-blockaded durations, as in Fig.~\ref{fig3}(d), each measurement at $B_{||} > 0$ can contain simultaneously information about the lifetimes of three states: the \Soo, \Tm\ and the \To.

Fig.~\ref{fig3}(f) shows the \Tm\ lifetime $\tau_T$ and \Soo-\To\ mixing times $\tau_+$ and $\tau_-$ calculated from the data in Fig.~\ref{fig3}(e). We find $\tau_+$ and $\tau_-$ from $\tau_u$ and $\tau_b^\prime$ using a rate equation model similar to the zero field case, but with no transitions to \Tp\ and \Tm\ included. This is because mixing from the \Soo\ or \To\ to the \Tp\ and \Tm\ will be suppressed due to their separation in energy. At $B_{||} \ge 0.5\units{T}$, the system spends so much time in the \Tm\ state that it is impractical to collect enough statistics to accurately determine $\tau_b^\prime$. Within the range of $B_{||}$ where $\tau_b^\prime$ can be measured, the \Soo-\To\ mixing rates are largely independent of field and similar to the rates seen at $B_{||} = 0$.

The time $\tau_T$ is the lifetime of the \Tm\ during the $\epsilon > 0$ half of the pulse and is well approximated as $\tau_T = \tau_b/2$ at high magnetic field. During the $\epsilon < 0$ half of the pulse, \Tm\ is the ground state and it will remain populated with high probability when $g\mu B_{||} > k_B T$. In the $\epsilon > 0$ half of the pulse the \Tm\ is the first excited state and can decay to the \Sot\ ground state at a of rate $\tau_T^{-1}$. Such transitions could be induced by phonons and a spin non-conserving process such as hyperfine coupling \cite{Johnson:2005p925,Coish:2005p125337,Taylor:2007p464} or spin-orbit coupling \cite{Tahan:2002p035314,Prada:2008p1187,Raith:2011p195318,Wang:2011p043716}. We find that the \Tm\ lifetime $\tau_T$ increases strongly with field, rising to 3 seconds by $B_{||} = 1\units{T}$.

In summary, we have shown that we can initialize the singlet-triplet qubit state into a singlet and subsequently measure, in single-shot mode, transitions to the (1,1) triplet states. Using this initialization and real-time measurement, we have measured the lifetime of singlet and triplet states versus magnetic field. At zero magnetic field, the lifetime for the singlet and all three triplets is $\sim 10\units{ms}$. At non-zero field, the \To\ and \Soo\ lifetimes are almost unchanged, whereas the \Tm\ lifetime grows significantly, reaching 3 seconds at 1T.

This work was supported by ARO and LPS (W911NF-08-1-0482) and by the United States Department of Defense. The US government requires publication of the following disclaimer: The views and conclusions contained in this document are those of the authors and should not be interpreted as representing the official policies, either expressly or implied, of the US Government.  This research utilized NSF-supported shared facilities at the University of Wisconsin-Madison. LV acknowledges financial support by a Starting Grant of the European Research Council (ERC) and by the Foundation for Fundamental Research on Matter (FOM).

\section{Supplemental information}
\newcounter{subfigure}
\renewcommand{\thefigure}{S\arabic{subfigure}}
\setcounter{subfigure}{1}

\subsection{Calibration of detuning energy}
\label{alpha}

We find the conversion between detuning voltage $V_\epsilon$ and detuning energy $\epsilon$ by measuring the width of the (1,1)-(0,2) transition as a function of fridge temperature $T_{fridge}$. The relevant temperature scale for the double-dot system is the electron temperature $T_e$, which we model as

\begin{equation}
\label{equ_Te}
T_e  = \sqrt{T_0^2 + T_{fridge}^2}
\end{equation}
where $T_0$ is the base electron temperature of our measurement setup. In the limit where $k_B T_e$ is greater than the inter-dot tunnel coupling $t_c$, the width of the (1,1)-(0,2) transition will be determined by $T_e$. When $k_B T_e \ll t_c$, the width is independent of temperature \cite{DiCarlo:2004p1440,Simmons:2009p3234}.

The line-shape of the inter-dot transition was measured by detecting the charge-sensor current $I_{QPC}$ with a lock-in amplifier while gates 2 and 4 were swept across the transition in the direction $V_\epsilon$ (as shown in Fig.~1(b) of the main text): $\Delta V_2 = -1.375 \Delta V_4$. The lock-in excitation was also applied in the $V_\epsilon$ direction by modulating gates 2 and 4 with a relative phase of 180 degrees and relative amplitudes of $V_{4p} = 0.35 V_{2p}$. The excitation frequency was $500\units{Hz}$ and the amplitude of the excitation at the gates was $\sim 20\units{\mu V}$ on gate 4. The results were fit with the expected line-shape in the thermally broadened limit, which is the derivative of a Fermi function:

\begin{equation}
\frac{dI_{QPC}}{dV_\epsilon} = A \mathrm{cosh}^{-2}\left( \frac{\alpha (V_\epsilon - V_{\epsilon 0})}{2 k_B T_e} \right) + C,
\end{equation}
where the voltage changes along the detuning axis are given by $V_4 = V_\epsilon$ and $V_2 = -V_\epsilon/1.375$, and $\alpha$ is the conversion factor between $V_\epsilon$ and detuning energy.

\begin{figure}
\includegraphics{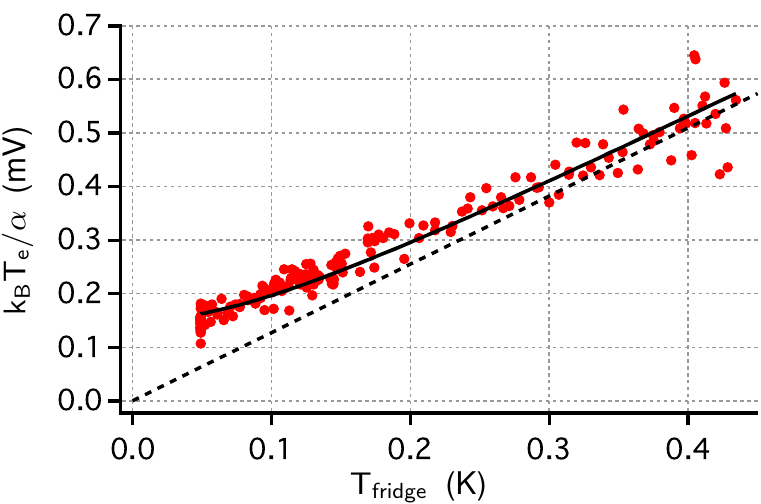}
\caption{\label{fig:tdep}Measured widths of the (1,1)-(0,2) transition as a function of the fridge temperature $T_{fridge}$.  The solid line is a fit of the form given by Equation~\ref{equ_Te} with $T_0 = 10.3\units{mK}$. The slope of the dependence at large temperatures yields the conversion between gate voltage along the detuning axis and energy: $\alpha = 67.6\units{meV/V}$. The dotted line shows the conversion if $T_0$ is neglected, which agrees well with the data at large temperatures.}
\end{figure}

Figure~\ref{fig:tdep} shows the measured values of the transition width $k_B T_e / \alpha$ as a function of $T_{fridge}$. We find that the width depends on temperature down $50\units{mK}$, confirming our assumption that we are in the limit $k_B T_e > t_c$, and also indicating that the base electron temperature $T_0$ is $< 50\units{mK}$. By fitting Equation~\ref{equ_Te} to the data, we find the calibration for the detuning energy to be $\alpha = 67.6\units{meV/V}$.

\subsection{Zero magnetic field rate equation model}

A set of coupled rate equations are used to extract \Soo-\To\ mixing rates at zero magnetic field from the measured values $\tau_u$ and $\tau_b$, which are the lifetimes of blockaded and un-blockaded configuration of the system. We relate $\tau_u$ and $\tau_b$ to the results of the rate equation model via two values: $P_{R} = e^{-T/\tau_u}$ is the probability of remaining un-blockaded in a single pulse cycle (the probability of returning to (0,2)), and $P_{N} = e^{-T/\tau_b}$ is the probability of remaining blockaded after a single pulse cycle (the probability of not returning to (0,2)). $T$ is the period of the pulse cycle, which in all the experiments consists of two halves of equal duration. For the first half of the cycle ($t < T/2$), the system is at negative detuning ($\epsilon < 0$), which favors the (1,1) state, while during the second half of the cycle ($T/2 < t < T$), the system is at positive detuning, where the (0,2) state is favored.

The physical mechanisms we consider are inter-dot tunneling between \Soo\ and \Sot, and singlet-triplet mixing due to a small field difference between the dots arising from the hyperfine interaction with background nuclear spins. At zero external field, the time-averaged effect of the small field difference between the dots is to cause all the (1,1) triplet states to mix with the (1,1) singlet at the same rate \cite{Coish:2005p125337,Taylor:2007p464}. This mixing rate will depend on the singlet-triplet splitting $J \approx t_c^2 / \epsilon$, where $t_c$ is the inter-dot tunnel coupling and $\epsilon$ is the detuning energy. The two halves of each pulse cycle will have different values of $|\epsilon|$ and, possibly, slightly different values of $t_c$. We therefore allow for two singlet-triplet mixing rates for the two halves of the cycle.

The double-dot states included in the model are \Soo, \Tm, \To, \Tp\ and \Sot, with occupation probabilities $p_1$, $p_2$, $p_3$, $p_4$ and $p_5$ respectively. During the first half of the pulse cycle, while the system is at negative detuning ($\epsilon < 0$), we allow all three triplet states to mix with \Soo\ at a rate $\rho_-$. The \Sot\ state is assumed to be un-populated because it will quickly tunnel to \Soo. The evolution of the occupation probabilities is given by:

\begin{eqnarray}
\label{eq_Soo1} \dot{p}_1(t) & = & \rho_- \left(p_2(t) + p_3(t) + p_4(t) - 3 p_1(t)\right)\\
\label{eq_Tm1} \dot{p}_2(t) & = & \rho_- \left(p_1(t) - p_2(t) \right)\\
\label{eq_To1} \dot{p}_3(t) & = & \rho_- \left(p_1(t) - p_3(t) \right)\\
\label{eq_Tp1} \dot{p}_4(t) & = & \rho_- \left(p_1(t) - p_4(t) \right)\\
\label{eq_Sot1} \dot{p}_5(t) & = & 0
\end{eqnarray}

During the second half of the pulse, from $t = T/2$ to $t = T$, the system is at positive detuning ($\epsilon > 0$) and the \Sot\ is now the ground state. The rate of singlet-triplet mixing is $\rho_+$, and we assume that all occupation of \Soo\ is instantly transferred to \Sot. The evolution of the occupation probabilities is given by:

\begin{eqnarray}
\label{eq_Soo2} \dot{p}_1(t) & = & 0 \\
\label{eq_Tm2} \dot{p}_2(t) & = & -  \rho_+ p_2(t) \\
\label{eq_To2} \dot{p}_3(t) & = & - \rho_+ p_3(t) \\
\label{eq_Tp2} \dot{p}_4(t) & = & - \rho_+ p_4(t) \\
\label{eq_Sot2} \dot{p}_5(t) & = & \rho_+ \left(p_2(t) + p_3(t) + p_4(t)\right)
\end{eqnarray}

We note that the model yields the same results if transitions between triplets are also allowed.

The two sets of coupled rate equations, Eqn.~\ref{eq_Soo1}-\ref{eq_Sot1} and Eqn.~\ref{eq_Soo2}-\ref{eq_Sot2}, are solved numerically to predict $P_N$ and $P_R$ for various $\rho_-$ and $\rho_+$. The values of $\rho_-$ and $\rho_+$ are then optimized to best match $P_N$ and $P_R$ given by the measurements. Errors on $\rho_-$ and $\rho_+$ are found by re-calculating the model multiple times while varying $P_N$ and $P_R$ within a range given by the errors on $\tau_b$ and $\tau_u$.

To calculate $P_R$, the initial conditions are set to unit probability of being in \Soo: $p_1(0) = 1$ and $p_{N\ne1}(0) = 0$. The occupation probabilities are evolved numerically for a time $T/2$ according to Eqn.~\ref{eq_Soo1}-\ref{eq_Sot1}. The resulting values of $p_N$ are then used as initial conditions for a further evolution for time $T/2$ according to Eqn.~\ref{eq_Soo2}-\ref{eq_Sot2}; however, all occupation of \Soo\ is first transferred to \Sot\ to simulate the fast inter-dot tunnel coupling of these states. The probability of remaining un-blockaded for a single pulse period is then given by $P_R = p_5(T)$.

$P_N$ is calculated in a similar way to $P_R$, but with the initial conditions being a unit probability of starting in a triplet state. The probability of remaining in a blockaded state for a single period is given by $P_N = \sum_{n=2}^4{p_n(T)}$. Because of the symmetry of the rate equations, the answer is independent of which triplet is initially occupied.

The lifetimes found by experiment at zero magnetic field are $\tau_u=23\pm3\units{ms}$ ($P_R = 0.87$) and $\tau_b=9.6\pm0.2\units{ms}$ ($P_N = 0.71$). We find that these values are consistent with \Soo-\To\ mixing times of $1/\rho_+ = 5.8\pm0.3 \units{ms}$ and $1/\rho_- = 24.5\pm3 \units{ms}$.


\begin{thebibliography}{23}
\expandafter\ifx\csname natexlab\endcsname\relax\def\natexlab#1{#1}\fi
\expandafter\ifx\csname bibnamefont\endcsname\relax
  \def\bibnamefont#1{#1}\fi
\expandafter\ifx\csname bibfnamefont\endcsname\relax
  \def\bibfnamefont#1{#1}\fi
\expandafter\ifx\csname citenamefont\endcsname\relax
  \def\citenamefont#1{#1}\fi
\expandafter\ifx\csname url\endcsname\relax
  \def\url#1{\texttt{#1}}\fi
\expandafter\ifx\csname urlprefix\endcsname\relax\def\urlprefix{URL }\fi
\providecommand{\bibinfo}[2]{#2}
\providecommand{\eprint}[2][]{\url{#2}}

\bibitem[{\citenamefont{Xiao et~al.}(2010)\citenamefont{Xiao, House, and
  Jiang}}]{Xiao:2010p096801}
\bibinfo{author}{\bibfnamefont{M.} \bibnamefont{Xiao}},
  \bibinfo{author}{\bibfnamefont{M.~G.} \bibnamefont{House}}, \bibnamefont{and}
  \bibinfo{author}{\bibfnamefont{H.~W.} \bibnamefont{Jiang}},
  \bibinfo{journal}{Phys. Rev. Lett.} \textbf{\bibinfo{volume}{104}},
  \bibinfo{pages}{096801} (\bibinfo{year}{2010}).

\bibitem[{\citenamefont{Morello et~al.}(2010)\citenamefont{Morello, Pla,
  Zwanenburg, Chan, Tan, Huebl, Mottonen, Nugroho, Yang, van Donkelaar
  et~al.}}]{Morello:2010p687}
\bibinfo{author}{\bibfnamefont{A.} \bibnamefont{Morello}},
  \bibnamefont{et~al.}, 
  \bibinfo{journal}{Nature (London)}
  \textbf{\bibinfo{volume}{467}}, \bibinfo{pages}{687} (\bibinfo{year}{2010}).

\bibitem[{\citenamefont{Simmons et~al.}(2011)\citenamefont{Simmons, Prance,
  Van~Bael, Koh, Shi, Savage, Lagally, Joynt, Friesen, Coppersmith
  et~al.}}]{Simmons:2011p156804}
\bibinfo{author}{\bibfnamefont{C.~B.} \bibnamefont{Simmons}},
  \bibinfo{author}{\bibfnamefont{J.~R.} \bibnamefont{Prance}},
  \bibinfo{author}{\bibfnamefont{B.~J.} \bibnamefont{Van~Bael}},
  \bibinfo{author}{\bibfnamefont{T.~S.} \bibnamefont{Koh}},
  \bibinfo{author}{\bibfnamefont{Z.}~\bibnamefont{Shi}},
  \bibinfo{author}{\bibfnamefont{D.~E.} \bibnamefont{Savage}},
  \bibinfo{author}{\bibfnamefont{M.~G.} \bibnamefont{Lagally}},
  \bibinfo{author}{\bibfnamefont{R.} \bibnamefont{Joynt}},
  \bibinfo{author}{\bibfnamefont{M.} \bibnamefont{Friesen}},
  \bibinfo{author}{\bibfnamefont{S.~N.} \bibnamefont{Coppersmith}},
  \bibnamefont{and} \bibinfo{author}{\bibfnamefont{M.~A.} \bibnamefont{Eriksson}},
  \bibinfo{journal}{Phys. Rev. Lett.}
  \textbf{\bibinfo{volume}{106}}, \bibinfo{pages}{156804}
  (\bibinfo{year}{2011}).

\bibitem[{\citenamefont{Hayes et~al.}()\citenamefont{Hayes, Kiselev, Borselli,
  Bui, Croke, Deelman, Maune, Milosavljevic, Moon, Ross
  et~al.}}]{Hayes:2009preprint}
\bibinfo{author}{\bibfnamefont{R.~R.} \bibnamefont{Hayes}},
  \bibnamefont{et~al.},  \bibinfo{journal}{e-print}  \bibinfo{note}{arXiv:0908.0173} (\bibinfo{year}{2009}).

\bibitem[{\citenamefont{Levy}(2002)}]{Levy:2002p1446}
\bibinfo{author}{\bibfnamefont{J.} \bibnamefont{Levy}}, \bibinfo{journal}{Phys.
  Rev. Lett.} \textbf{\bibinfo{volume}{89}}, \bibinfo{pages}{147902}
  (\bibinfo{year}{2002}).

\bibitem[{\citenamefont{Petta et~al.}(2005{\natexlab{a}})\citenamefont{Petta,
  Johnson, Taylor, Laird, Yacoby, Lukin, Marcus, Hanson, and
  Gossard}}]{Petta:2005p2180}
\bibinfo{author}{\bibfnamefont{J.~R.} \bibnamefont{Petta}},
  \bibinfo{author}{\bibfnamefont{A.~C.} \bibnamefont{Johnson}},
  \bibinfo{author}{\bibfnamefont{J.~M.} \bibnamefont{Taylor}},
  \bibinfo{author}{\bibfnamefont{E.~A.} \bibnamefont{Laird}},
  \bibinfo{author}{\bibfnamefont{A.}~\bibnamefont{Yacoby}},
  \bibinfo{author}{\bibfnamefont{M.~D.} \bibnamefont{Lukin}},
  \bibinfo{author}{\bibfnamefont{C.~M.} \bibnamefont{Marcus}},
  \bibinfo{author}{\bibfnamefont{M.~P.} \bibnamefont{Hanson}},
  \bibnamefont{and} \bibinfo{author}{\bibfnamefont{A.~C.}
  \bibnamefont{Gossard}}, \bibinfo{journal}{Science}
  \textbf{\bibinfo{volume}{309}}, \bibinfo{pages}{2180}
  (\bibinfo{year}{2005}{\natexlab{a}}).

\bibitem[{\citenamefont{Foletti et~al.}(2009)\citenamefont{Foletti, Bluhm,
  Mahalu, Umansky, and Yacoby}}]{Foletti:2009p903}
\bibinfo{author}{\bibfnamefont{S.} \bibnamefont{Foletti}},
  \bibinfo{author}{\bibfnamefont{H.} \bibnamefont{Bluhm}},
  \bibinfo{author}{\bibfnamefont{D.} \bibnamefont{Mahalu}},
  \bibinfo{author}{\bibfnamefont{V.} \bibnamefont{Umansky}}, \bibnamefont{and}
  \bibinfo{author}{\bibfnamefont{A.} \bibnamefont{Yacoby}},
  \bibinfo{journal}{Nature Phys.} \textbf{\bibinfo{volume}{5}},
  \bibinfo{pages}{903} (\bibinfo{year}{2009}).

\bibitem[{\citenamefont{Johnson et~al.}(2005)\citenamefont{Johnson, Petta,
  Taylor, Yacoby, Lukin, Marcus, Hanson, and Gossard}}]{Johnson:2005p925}
\bibinfo{author}{\bibfnamefont{A.~C.} \bibnamefont{Johnson}},
  \bibinfo{author}{\bibfnamefont{J.~R.} \bibnamefont{Petta}},
  \bibinfo{author}{\bibfnamefont{J.~M.} \bibnamefont{Taylor}},
  \bibinfo{author}{\bibfnamefont{A.} \bibnamefont{Yacoby}},
  \bibinfo{author}{\bibfnamefont{M.~D.} \bibnamefont{Lukin}},
  \bibinfo{author}{\bibfnamefont{C.~M.} \bibnamefont{Marcus}},
  \bibinfo{author}{\bibfnamefont{M.~P.} \bibnamefont{Hanson}},
  \bibnamefont{and} \bibinfo{author}{\bibfnamefont{A.~C.}
  \bibnamefont{Gossard}}, \bibinfo{journal}{Nature (London)}
  \textbf{\bibinfo{volume}{435}}, \bibinfo{pages}{925} (\bibinfo{year}{2005}).

\bibitem[{\citenamefont{Petta et~al.}(2005{\natexlab{b}})\citenamefont{Petta,
  Johnson, Yacoby, Marcus, Hanson, and Gossard}}]{Petta:2005p161301}
\bibinfo{author}{\bibfnamefont{J.~R.} \bibnamefont{Petta}},
  \bibinfo{author}{\bibfnamefont{A.~C.} \bibnamefont{Johnson}},
  \bibinfo{author}{\bibfnamefont{A.} \bibnamefont{Yacoby}},
  \bibinfo{author}{\bibfnamefont{C.~M.} \bibnamefont{Marcus}},
  \bibinfo{author}{\bibfnamefont{M.~P.} \bibnamefont{Hanson}},
  \bibnamefont{and} \bibinfo{author}{\bibfnamefont{A.~C.} \bibnamefont{Gossard}}, 
  \bibinfo{journal}{Phys. Rev. B}
  \textbf{\bibinfo{volume}{72}}, \bibinfo{pages}{161301}
  (\bibinfo{year}{2005}{\natexlab{b}}).

\bibitem[{\citenamefont{Simmons et~al.}(2009)\citenamefont{Simmons, Thalakulam,
  Rosemeyer, Bael, Sackmann, Savage, Lagally, Joynt, Friesen, Coppersmith
  et~al.}}]{Simmons:2009p3234}
\bibinfo{author}{\bibfnamefont{C.~B.} \bibnamefont{Simmons}},
  \bibinfo{author}{\bibfnamefont{M.} \bibnamefont{Thalakulam}},
  \bibinfo{author}{\bibfnamefont{B.~M.} \bibnamefont{Rosemeyer}},
  \bibinfo{author}{\bibfnamefont{B.~J.} \bibnamefont{Van Bael}},
  \bibinfo{author}{\bibfnamefont{E.~K.} \bibnamefont{Sackmann}},
  \bibinfo{author}{\bibfnamefont{D.~E.} \bibnamefont{Savage}},
  \bibinfo{author}{\bibfnamefont{M.~G.} \bibnamefont{Lagally}},
  \bibinfo{author}{\bibfnamefont{R.} \bibnamefont{Joynt}},
  \bibinfo{author}{\bibfnamefont{M.} \bibnamefont{Friesen}},
  \bibinfo{author}{\bibfnamefont{S.~N.} \bibnamefont{Coppersmith}},
  \bibnamefont{and} \bibinfo{author}{\bibfnamefont{M.~A.} \bibnamefont{Eriksson}},
  \bibinfo{journal}{Nano Lett.}
  \textbf{\bibinfo{volume}{9}}, \bibinfo{pages}{3234} (\bibinfo{year}{2009}).

\bibitem[{\citenamefont{Hanson et~al.}(2007)\citenamefont{Hanson, Kouwenhoven,
  Petta, Tarucha, and Vandersypen}}]{Hanson:2007p1217}
\bibinfo{author}{\bibfnamefont{R.} \bibnamefont{Hanson}},
  \bibinfo{author}{\bibfnamefont{L.~P.} \bibnamefont{Kouwenhoven}},
  \bibinfo{author}{\bibfnamefont{J.~R.} \bibnamefont{Petta}},
  \bibinfo{author}{\bibfnamefont{S.} \bibnamefont{Tarucha}}, \bibnamefont{and}
  \bibinfo{author}{\bibfnamefont{L.~M.~K.} \bibnamefont{Vandersypen}},
  \bibinfo{journal}{Rev. Mod. Phys.} \textbf{\bibinfo{volume}{79}},
  \bibinfo{pages}{1217} (\bibinfo{year}{2007}).

\bibitem[{\citenamefont{Shaji et~al.}(2008)\citenamefont{Shaji, Simmons,
  Thalakulam, Klein, Qin, Luo, Savage, Lagally, Rimberg, Joynt
  et~al.}}]{Shaji:2008p540}
\bibinfo{author}{\bibfnamefont{N.} \bibnamefont{Shaji}},
  \bibnamefont{et~al.}, \bibinfo{journal}{Nature Phys.}
  \textbf{\bibinfo{volume}{4}}, \bibinfo{pages}{540} (\bibinfo{year}{2008}).

\bibitem[{\citenamefont{Borselli et~al.}(2011)\citenamefont{Borselli, Eng,
  Croke, Maune, Huang, Ross, Kiselev, Deelman, Alvarado-Rodriguez, Schmitz
  et~al.}}]{Borselli:2011p063109}
\bibinfo{author}{\bibfnamefont{M.~G.} \bibnamefont{Borselli}},
  \bibnamefont{et~al.}, \bibinfo{journal}{Appl. Phys. Lett.}
  \textbf{\bibinfo{volume}{99}}, \bibinfo{eid}{063109}
  (\bibinfo{year}{2011}).

\bibitem[{\citenamefont{Lai et~al.}(2010)\citenamefont{Lai, Lim, Yang,
  Zwanenburg, Coish, Qassemi, Morello, and Dzurak}}]{Lai:2010preprint}
\bibinfo{author}{\bibfnamefont{N.~S.} \bibnamefont{Lai}},
  \bibinfo{author}{\bibfnamefont{W.~H.} \bibnamefont{Lim}},
  \bibinfo{author}{\bibfnamefont{C.~H.} \bibnamefont{Yang}},
  \bibinfo{author}{\bibfnamefont{F.~A.} \bibnamefont{Zwanenburg}},
  \bibinfo{author}{\bibfnamefont{W.~A.} \bibnamefont{Coish}},
  \bibinfo{author}{\bibfnamefont{F.} \bibnamefont{Qassemi}},
  \bibinfo{author}{\bibfnamefont{A.} \bibnamefont{Morello}}, \bibnamefont{and}
  \bibinfo{author}{\bibfnamefont{A.~S.} \bibnamefont{Dzurak}},
  \bibinfo{journal}{e-print}  \bibinfo{note}{arXiv:1012.1410} (\bibinfo{year}{2010}).

\bibitem[{\citenamefont{Coish and Loss}(2005)}]{Coish:2005p125337}
\bibinfo{author}{\bibfnamefont{W.~A.} \bibnamefont{Coish}} \bibnamefont{and}
  \bibinfo{author}{\bibfnamefont{D.}~\bibnamefont{Loss}},
  \bibinfo{journal}{Phys. Rev. B} \textbf{\bibinfo{volume}{72}},
  \bibinfo{pages}{125337} (\bibinfo{year}{2005}).

\bibitem[{\citenamefont{Taylor et~al.}(2007)\citenamefont{Taylor, Petta,
  Johnson, Yacoby, Marcus, and Lukin}}]{Taylor:2007p464}
\bibinfo{author}{\bibfnamefont{J.~M.} \bibnamefont{Taylor}},
  \bibinfo{author}{\bibfnamefont{J.~R.} \bibnamefont{Petta}},
  \bibinfo{author}{\bibfnamefont{A.~C.} \bibnamefont{Johnson}},
  \bibinfo{author}{\bibfnamefont{A.}~\bibnamefont{Yacoby}},
  \bibinfo{author}{\bibfnamefont{C.~M.} \bibnamefont{Marcus}},
  \bibnamefont{and} \bibinfo{author}{\bibfnamefont{M.~D.} \bibnamefont{Lukin}},
  \bibinfo{journal}{Phys. Rev. B} \textbf{\bibinfo{volume}{76}},
  \bibinfo{pages}{035315} (\bibinfo{year}{2007}).

\bibitem[{\citenamefont{Assali et~al.}(2011)\citenamefont{Assali, Petrilli,
  Capaz, Koiller, Hu, and Das~Sarma}}]{Assali:2011p165301}
\bibinfo{author}{\bibfnamefont{L.~V.~C.} \bibnamefont{Assali}},
  \bibinfo{author}{\bibfnamefont{H.~M.} \bibnamefont{Petrilli}},
  \bibinfo{author}{\bibfnamefont{R.~B.} \bibnamefont{Capaz}},
  \bibinfo{author}{\bibfnamefont{B.} \bibnamefont{Koiller}},
  \bibinfo{author}{\bibfnamefont{X.} \bibnamefont{Hu}}, \bibnamefont{and}
  \bibinfo{author}{\bibfnamefont{S.} \bibnamefont{Das~Sarma}},
  \bibinfo{journal}{Phys. Rev. B} \textbf{\bibinfo{volume}{83}},
  \bibinfo{pages}{165301} (\bibinfo{year}{2011}).

\bibitem[{\citenamefont{Koppens et~al.}(2005)\citenamefont{Koppens, Folk,
  Elzerman, Hanson, van Beveren, Vink, Tranitz, Wegscheider, Kouwenhoven, and
  Vandersypen}}]{Koppens:2005p717}
\bibinfo{author}{\bibfnamefont{F.~H.~L.} \bibnamefont{Koppens}},
  \bibinfo{author}{\bibfnamefont{J.~A.} \bibnamefont{Folk}},
  \bibinfo{author}{\bibfnamefont{J.~M.} \bibnamefont{Elzerman}},
  \bibinfo{author}{\bibfnamefont{R.} \bibnamefont{Hanson}},
  \bibinfo{author}{\bibfnamefont{L.~H.} \bibnamefont{Willems van Beveren}},
  \bibinfo{author}{\bibfnamefont{I.~T.} \bibnamefont{Vink}},
  \bibinfo{author}{\bibfnamefont{H.~P.} \bibnamefont{Tranitz}},
  \bibinfo{author}{\bibfnamefont{W.} \bibnamefont{Wegscheider}},
  \bibinfo{author}{\bibfnamefont{L.~P.} \bibnamefont{Kouwenhoven}},
  \bibnamefont{and}
  \bibinfo{author}{\bibfnamefont{L.~M.~K.} \bibnamefont{Vandersypen}},
  \bibinfo{journal}{Science} \textbf{\bibinfo{volume}{309}},
  \bibinfo{pages}{1346} (\bibinfo{year}{2005}).

\bibitem[{\citenamefont{Tahan et~al.}(2002)\citenamefont{Tahan, Friesen, and
  Joynt}}]{Tahan:2002p035314}
\bibinfo{author}{\bibfnamefont{C.} \bibnamefont{Tahan}},
  \bibinfo{author}{\bibfnamefont{M.} \bibnamefont{Friesen}}, \bibnamefont{and}
  \bibinfo{author}{\bibfnamefont{R.} \bibnamefont{Joynt}},
  \bibinfo{journal}{Phys. Rev. B} \textbf{\bibinfo{volume}{66}},
  \bibinfo{pages}{035314} (\bibinfo{year}{2002}).

\bibitem[{\citenamefont{Prada et~al.}(2008)\citenamefont{Prada, Blick, and
  Joynt}}]{Prada:2008p1187}
\bibinfo{author}{\bibfnamefont{M.} \bibnamefont{Prada}},
  \bibinfo{author}{\bibfnamefont{R.~H.} \bibnamefont{Blick}}, \bibnamefont{and}
  \bibinfo{author}{\bibfnamefont{R.} \bibnamefont{Joynt}},
  \bibinfo{journal}{Phys. Rev. B} \textbf{\bibinfo{volume}{77}},
  \bibinfo{pages}{115438}  (\bibinfo{year}{2008}).

\bibitem[{\citenamefont{Raith et~al.}(2011)\citenamefont{Raith, Stano, and
  Fabian}}]{Raith:2011p195318}
\bibinfo{author}{\bibfnamefont{M.} \bibnamefont{Raith}},
  \bibinfo{author}{\bibfnamefont{P.} \bibnamefont{Stano}}, \bibnamefont{and}
  \bibinfo{author}{\bibfnamefont{J.} \bibnamefont{Fabian}},
  \bibinfo{journal}{Phys. Rev. B} \textbf{\bibinfo{volume}{83}},
  \bibinfo{pages}{195318} (\bibinfo{year}{2011}).

\bibitem[{\citenamefont{Wang and Wu}(2011)}]{Wang:2011p043716}
\bibinfo{author}{\bibfnamefont{L.} \bibnamefont{Wang}} \bibnamefont{and}
  \bibinfo{author}{\bibfnamefont{M.~W.} \bibnamefont{Wu}},
  \bibinfo{journal}{J. App. Phys.} \textbf{\bibinfo{volume}{110}},
  \bibinfo{pages}{043716} (\bibinfo{year}{2011}).
  
  \bibitem[{\citenamefont{Dicarlo et~al.}(2004)\citenamefont{Dicarlo, Lynch,
  Johnson, Childress, Crockett, and Marcus}}]{DiCarlo:2004p1440}
\bibinfo{author}{\bibfnamefont{L.}~\bibnamefont{Dicarlo}},
  \bibinfo{author}{\bibfnamefont{H.~J.} \bibnamefont{Lynch}},
  \bibinfo{author}{\bibfnamefont{A.~C.} \bibnamefont{Johnson}},
  \bibinfo{author}{\bibfnamefont{L.~I.} \bibnamefont{Childress}},
  \bibinfo{author}{\bibfnamefont{K.} \bibnamefont{Crockett}}, \bibnamefont{and}
  \bibinfo{author}{\bibfnamefont{C.~M.} \bibnamefont{Marcus}},
  \bibinfo{journal}{Phys. Rev. Lett.} \textbf{\bibinfo{volume}{92}},
  \bibinfo{pages}{226801} (\bibinfo{year}{2004}).

\end{thebibliography}
\end{document}